\documentclass[11pt]{article}%
\usepackage{algorithm}
\usepackage{algpseudocode}
\usepackage{amssymb}
\usepackage{amsfonts}
\usepackage{amsmath}
\usepackage{amsthm}
\usepackage{hyperref}
\usepackage{graphicx}
\usepackage{multirow}
\usepackage{booktabs}
\usepackage{bm}
\usepackage{relsize}
\providecommand{\U}[1]{\protect\rule{.1in}{.1in}}
\usepackage{tikz}
\usetikzlibrary{shapes}
\usepackage{pgfplots}
\usepackage{verbatim}
\usepgfplotslibrary{statistics}
\pgfplotsset{compat=1.8}
\renewcommand\and{\end{tabular}\kern-\tabcolsep\ and\ \kern-\tabcolsep\begin{tabular}[t]{c}}
\let\origthanks\thanks
\renewcommand\thanks[1]{\begingroup\let\rlap\relax\origthanks{#1}\endgroup}
\usepackage{autonum}

\usepackage{mathtools}
\mathtoolsset{showonlyrefs}

\makeatletter
\newcounter{phase}[algorithm]
\newlength{\phaserulewidth}

\makeatother

\usetikzlibrary{positioning}
\newif\ifOLDSTUFF
\OLDSTUFFfalse

\newcommand{\argmax}{\operatornamewithlimits{argmax}}

\providecommand{\keywords}[1]
{
  \small	
  \textbf{\textit{Keywords---}} #1
}

\begin{document}

\title{Two-population SIR model and strategies to reduce mortality in pandemics}
\author{Long Ma, Maksim Kitsak and Piet Van Mieghem}
\date{Delft University of Technology\\
}
\maketitle

\begin{abstract}
Despite many studies on the transmission mechanism of the Severe acute respiratory syndrome coronavirus 2 (SARS-CoV-2), it remains still challenging to efficiently reduce mortality. In this work, we apply a two-population Susceptible-Infected-Removed (SIR) model to investigate the COVID-19 spreading when contacts between elderly and non-elderly individuals are reduced due to the high mortality risk of elderly people. We discover that the reduction of connections between two populations can delay the death curve
but cannot well reduce the final mortality. We propose a merged SIR model, which advises elderly individuals to interact less with their non-elderly connections at the initial stage but interact more with their non-elderly relationships later, to reduce the final mortality. Finally, immunizing elderly hub individuals can also significantly decrease mortality.
\end{abstract}


\section{Introduction}

In many countries, the first wave of the Coronavirus disease 2019 (COVID-19) appeared in early 2020. In the summer of 2020, the spread of COVID-19 was significantly reduced due to strict restrictions \cite{jordan2020covid} and weather effects \cite{merow2020seasonality}. At that time, the majority of the population and politicians were hoping for the end of the COVID-19 pandemic. At the beginning of autumn, students went back to school, which marked the beginning of the second wave. However, rising infections were not taken seriously because infections were mainly among the young population and no significant hospitalization and deaths were observed \cite{Reimann2020}. Simultaneously, the high decease rate and self-preservation have caused that many elderly individuals reduced their contact with young people  \cite{armitage2020covid}. In October 2020, the hospitalization rates in many countries started increasing and the second COVID-19 wave was born. At the beginning of 2021, more contagious mutations of the coronavirus marked the third wave in many countries  \cite{leung2021early}. Even though there are COVID-19 vaccines, the distribution in many countries is painfully slow. Moreover, SARS-CoV-2 viral mutations lead to uncertainty about the effectiveness of recent vaccines. The third wave might not be the last COVID-19 pandemic and efficient strategies to reduce mortality will remain on the agenda.

The Susceptible-Infected-Removed (SIR) model \cite{fernandez2020estimating,morris2021optimal} and its variations are commonly applied to describe the COVID-19 pandemic \cite{fahira2021effect,cooper2020sir,pastor2015epidemic,faranda2020modeling,ceylan2020estimation} and to forecast the number of infected and deceased cases in a population \cite{prasse2020network,achterberg2020comparing,neves2020predicting,yang2020modified,zhou2020forecasting,massonnaud2020covid}.  
The ratio of new deceased elderly cases to new deceased non-elderly cases each day is expected to be constant over time in classic epidemic models but is time-varying in reality. Recent works start to consider the age-structured SIR model to describe the COVID-19 pandemic more realistic \cite{arenas2020mathematical,radulescu2020management,di2020impact,prem2020effect,zhao2020five}. The age-structured SIR model divides the whole population into several age groups and the infection rates are age-dependent. Real data reveal that the elderly infected had a 30- to 100- fold higher risk of dying than younger individuals in many European countries \cite{ioannidis2020population}. Here, elderly and non-elderly individuals are respectively defined as individuals who are $ <65 $ years old and $ \geq 65 $ years old \cite{ioannidis2020population}. The elderly population accounts for a proportion of around $ 20\% $ in many European countries \cite{worldbank2019population}. Since the main difference in the COVID-19 pandemic is between elderly and non-elderly individuals, we construct a two-population SIR model \cite{magal2016final} as follows:
\begin{enumerate}
\item There are two sub-populations: non-elderly and elderly individuals uniformly distributed over the social contact network. The virus spreading in a region is likely to start from non-elderly individuals because the virus can be carried into a community from other areas by commuters \cite{pei2018forecasting} and most commuters are non-elderly individuals.
\item There are four infection rates between and within non-elderly and elderly individuals. We believe that the highest infection rate is among elderly people. Elderly individuals are advised to a kind of self-isolation to protect themselves \cite{hajek2021social}. Staying in relative isolation from non-elderly people could be feasible, but some strong connections among elderly individuals, e.g., couples and people in the same nursing home, cannot be cut off. Conversely, the ties among elderly individuals will be stronger when their connections with non-elderly individuals are significantly reduced. The second highest infection rate is among the non-elderly population. The inter-group infection rates are the smallest since elderly individuals are afraid of being infected by non-elderly individuals. The infection rates between non-elderly and elderly individuals are low, but not zero, as elderly people still depend on younger people one way or another.
\end{enumerate}

This article first investigates the features of fatality curves in the two-population SIR model when the connection between two populations is reduced. It shows that non-elderly deceased cases are prone to occur at the initial stage and most elderly deceased patients appear more often at a later stage. The difference in infection probability between non-elderly and elderly individuals is significant when the inter-population infection rates are low and the infection rate among elderly individuals is slightly above the epidemic threshold. The final mortality, however, cannot be reduced by only limiting the connection between two populations. Moreover, reducing the infection rate among non-elderly individuals, e.g., closing schools, can also not efficiently reduce mortality. In this work, we propose a merged SIR model to reduce the final mortality significantly. There are two stages in the merged SIR model: in the first stage, the model is the same as the two-population SIR model of Magal {\em et al.} \cite{magal2016final} and in the second stage, the merged SIR model reduces to the standard SIR model. The physical meaning of the merged SIR model is that elder people are advised to reduce their connections with non-elderly individuals at the beginning of the pandemic and interact more with non-elderly individuals later. The merged SIR model benefits the mortality reduction since many recovered non-elderly people can protect the susceptible elderly individuals.

Compartmental epidemic models assume that social contact networks are homogeneous with an infinite network size, but the actual network size is finite and the degree distributions of many real social networks follow a power law \cite{eubank2004structural} with an exponent $ \gamma\in[2,3] $. We thus simulate the two-population SIR model on a scale-free network with a realistic network size to investigate the effect of network topology on the reduction of mortality. By comparing the simulation results of the two-population SIR model for the scale-free network and the Erdős–Rényi random network \cite{erdHos1960evolution}, the epidemic spreading in the heterogeneous network is much faster due to the star (or super spreader) effect. The reduction of connections between elderly and non-elderly individuals cannot decrease mortality in the compartmental epidemic model, but can reduce the mortality in the two-population SIR epidemic on the scale-free network. The merged SIR model is the best strategy to efficiently mitigate the mortality. Finally, we illustrate that mortality can be efficiently reduced by only immunizing rare elderly hub individuals.

\section{Two-population SIR model}

The two-population SIR model was first proposed by Magal \textit{et al.} \cite{magal2016final}. Similar models, that incorporate the underlying contact graph, are the networked SIR model proposed by Youssef and Scoglio \cite{youssef2011individual}, that was later entirely generalized to GEMF in \cite{sahneh2013generalized}. Our work here applies the two-population SIR model to a realistic scenario related to the COVID-19 pandemic, systematically analyzes the death-related curve features, explores the effect of restrictions on mortality reduction and proposes an improved model to reduce the final mortality.

\begin{figure}[!hbtp]
\begin{center}
\includegraphics[
width=0.4\linewidth
]%
{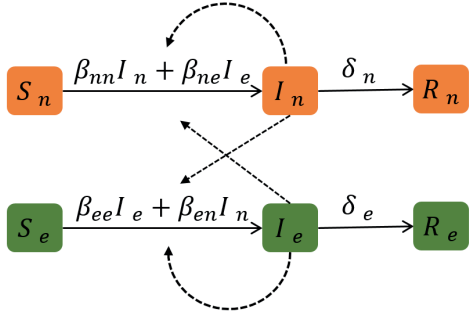}%
\caption{Schematic depiction of the two-population SIR model. There are two populations in the model: non-elderly individuals (highlighted in orange) and elderly individuals (highlighted in green). There are four different infection rates $ \beta_{nn} $, $ \beta_{ne} $, $ \beta_{en} $ and $ \beta_{ee} $ between and among the populations and two different removed rates $ \delta_n $ and $ \delta_e $ for each population.}%
\label{fig_diagram}%
\end{center}
\end{figure}

Suppose that the elderly and non-elderly populations are well-mixed and large enough, then the fractions of susceptible individuals $ S(t) $, infectious individuals $ I(t) $ and removed (recovered or deceased) individuals $ R(t) $ at time $t$ are reasonably well modeled by the following well-known differential equations:
\begin{equation}
\left\{
\begin{aligned}
&\dfrac{dS(t)}{dt} = -\text{diag}(S(t))BI(t),\\
&\dfrac{dI(t)}{dt} = \text{diag}(S(t))BI(t)-EI(t),\\
&\dfrac{dR(t)}{dt} = EI(t),
\end{aligned}
\right.
\label{eqn:well-mixed_sir_model}
\end{equation}
where the vectors of fractions $ S(t) $, $ I(t) $ and $ R(t) $ are respectively,
\begin{equation}
    S(t)=\begin{pmatrix}
  S_{n}(t) \\
  S_{e}(t) \\
\end{pmatrix}, I(t)=\begin{pmatrix}
  I_{n}(t) \\
  I_{e}(t) \\
\end{pmatrix}, R(t)=\begin{pmatrix}
  R_{n}(t) \\
  R_{e}(t) \\
\end{pmatrix}, 
\end{equation}
and the matrices $E$ (removed rates) and $B$ (infection rates) are respectively
\begin{equation}
    E=\begin{pmatrix}
  \delta_n & 0\\
  0 & \delta_e\\
  \end{pmatrix},
      B=\begin{pmatrix}
  \beta_{nn} & \beta_{ne}\\
  \beta_{en} & \beta_{ee}\\
  \end{pmatrix},
\end{equation}
where $ \beta_{ne} $ denotes the infection rate from elderly infectious individuals to non-elderly susceptible individuals, $ \beta_{en} $ denotes the infection rate from non-elderly infectious individuals to elderly susceptible individuals, $ \beta_{nn} $ denotes the infection rate among non-elderly individuals, $ \beta_{ee} $ denotes the infection rate among elderly individuals, $ \delta_n $ denotes the removed rate for non-elderly infectious individuals and $\delta_e$ denotes the removed rate for elderly infectious individuals. To simplify, we let the infection rates between two populations be equal, $ \beta_{ne} = \beta_{en} =\epsilon \beta_{nn}$, and thus the matrix $B$ can be rewritten as
\begin{equation}\label{B_matrix}
    B=\beta_{nn}
    \begin{pmatrix}
  1 & \epsilon\\
  \epsilon & \kappa\\
  \end{pmatrix}.
\end{equation}
For the COVID-19 pandemic, it holds that $\epsilon\ll 1$ and $\kappa\geq 1$. Furthermore, we have that the non-elderly fractions $ S_{n}(t) + I_{n}(t) + R_{n}(t) = N_n $ and the elderly fractions $ {S}_{e}(t) + {I}_{e}(t) + R_{e}(t) = N_e $, where $ N_n $ denotes the fraction of non-elderly population and $ N_e $ denotes the fraction of elderly population. The two-population SIR model assumes that the total population is unchanged and thus $ N_n+N_e=1 $. We denote the initial state by $v[0] = (S_n[0], I_n[0], R_{n}[0], S_e[0], I_e[0], R_{e}[0])$. A schematic depiction of the two-population SIR model is shown in Fig.~\ref{fig_diagram}. The infectious individuals will turn to be immune with a recovery rate ($\xi_n$ for non-elderly individuals and $\xi_e$ for elderly individuals) or deceased with a fatality rate ($ \eta_n $ for non-elderly individuals and $ \eta_e $  for elderly individuals). It holds that the removed rates $ \delta_n = \eta_n + \xi_n $ and $ \delta_e = \eta_e + \xi_e $. This work focuses on the fractions of \textit{new} deceased non-elderly and elderly cases that are $\boldsymbol{\eta_n{I}_{n}(t)}$ and $ \boldsymbol{\eta_e{I}_{e}(t)}$, respectively. We are also interested in the fractions of deceased non-elderly and elderly cases that are $ \boldsymbol{D_n(t)= \eta_nR_n(t)/\delta_n }$ and $\boldsymbol{D_e(t)= \eta_eR_e(t)/\delta_e} $.

By numerical solving Equations \eqref{eqn:well-mixed_sir_model}, we analyze the effect of infection rates on the following four death-related curve features,
\begin{enumerate}
    \item maximum of $\eta_n{I}_{n}(t)$ and $ \eta_e{I}_{e}(t)$: $\max\limits_{t\geq 0} \eta_n{I}_{n}(t)$ and $\max\limits_{t\geq 0} \eta_e{I}_{e}(t)$,
    \item time points at which the maximum of $\eta_n{I}_{n}(t)$ and $ \eta_e{I}_{e}(t)$ occur: $\argmax\limits_{t\geq 0} \eta_n{I}_{n}(t)$ and $\argmax\limits_{t\geq 0} \eta_e{I}_{e}(t)$,
    \item time difference between two arguments of the maxima: $\argmax\limits_{t\geq 0} \eta_e{I}_{e}(t)-\argmax\limits_{t\geq 0} \eta_n{I}_{n}(t)$,
    \item fractions of final deceased non-elderly cases $D_n(\infty)$ and elderly cases $D_e(\infty)$.
\end{enumerate}
In this work, we set the fraction of non-elderly individuals as $ N_n=0.8 $, the fraction of elderly individuals as $ N_e=0.2 $ and the removed rates as $ \delta_n=\delta_e=0.1 $. The fatality rates for non-elderly and elderly infections are set to be $ \eta_n=0.0001 $ and $ \eta_e=0.01 $, respectively. The initial state is set as $ v[0]=(0.7999,0.0001,0,0.2,0,0) $. These parameters are set based on real data. The elderly population makes up around $ 20\% $ of the whole population in many European countries \cite{worldbank2019population}. Elderly people who were infected had 30- to 100- fold higher risk of dying than younger people in several European countries \cite{ioannidis2020population}. The time to recovery or death is on average around 10 days \cite{centers2020clinical}.
We also investigate various parameter settings and find that the changing of these parameters has no much effect on the main conclusions drawn in this paper.

\begin{figure}[!hbtp]
\begin{center}
\includegraphics[
width=0.8\linewidth
]%
{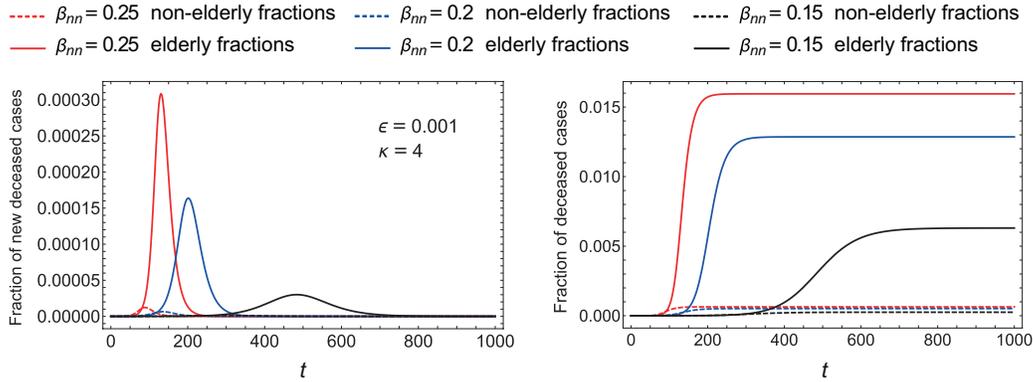}%
\caption{Death-related curves of the two-population SIR model with different infection rates $\beta_{nn}$. The left figure is about the fractions of new deceased cases $ \eta_nI_n(t) $ for non-elderly individuals (dashed curves) and $ \eta_eI_e(t) $ for elderly individuals (solid curves). The right figure is about the fractions of deceased cases $D_n(t)$ for non-elderly individuals (dashed curves) and $D_e(t)$ for elderly individuals (solid curves). Parameters $ \epsilon $ and $\kappa$ are set as $ 0.001 $ and $4$, respectively. Three different infection rates $\beta_{nn}$ are considered: $\beta_{nn}=0.25$ (red curves), $\beta_{nn}=0.2$ (blue curves) and $\beta_{nn}=0.15$ (black curves).}%
\label{fig_numerical_solving}%
\end{center}
\end{figure}

There are three parameters in matrix \eqref{B_matrix}, which are $\beta_{nn}$, $\epsilon$ and $\kappa$. We first set $ \epsilon = 0.001 $ and $\kappa=4$ and study the effect of the infection rate $\beta_{nn}$ on death-related curves. Figure \ref{fig_numerical_solving} reveals that both the non-elderly related curves and elderly related curves are significantly affected by the parameter $\beta_{nn}$. The time difference $\argmax\limits_{t\geq 0} \eta_e{I}_{e}(t)-\argmax\limits_{t\geq 0} \eta_n{I}_{n}(t)$ is positive and increases with the infection rate $\beta_{nn}$ deceasing. The final non-elderly deceased fraction $D_n(\infty)$ and elderly deceased fraction $D_e(\infty)$ increase with the infection rate $\beta_{nn}$.

\begin{figure}[!hbtp]
\begin{center}
\includegraphics[
width=\linewidth
]%
{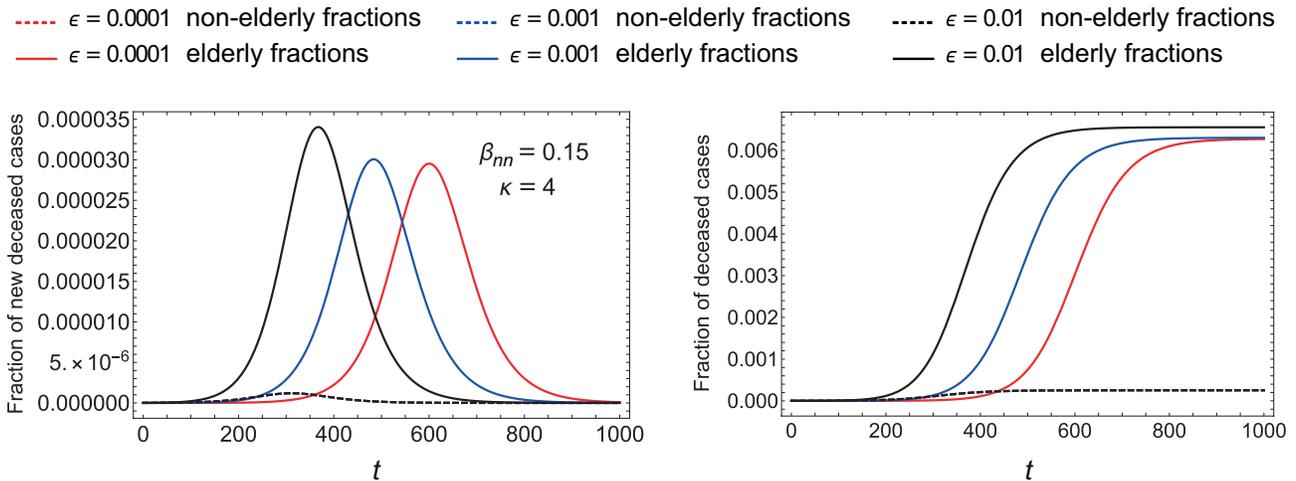}%
\caption{Death-related curves of the two-population SIR model with different parameters $\epsilon$. Parameters $ \beta_{nn}$ and $\kappa$ are set as $ 0.15 $ and $4$, respectively. Three different parameters $\epsilon$ are considered: $\epsilon=0.0001$ (red curves), $\epsilon=0.001$ (blue curves) and $\epsilon=0.01$ (black curves).}
\label{fig_2}%
\end{center}
\end{figure}

We further set parameters $ \beta_{nn} = 0.15 $ and $\kappa=4$ and study the effect of parameter $\epsilon$ on death-related curves. Figure \ref{fig_2} shows the death-related fractions with different parameter $\epsilon$. It indicates that the parameter $\epsilon$ has almost no effect on non-elderly related curves. The final deceased fractions are little affected by the parameter $\epsilon$. The effect of smaller $\epsilon$ is approximately to delay the elderly related curves and there will be larger time difference $\argmax\limits_{t\geq 0} \eta_e{I}_{e}(t)-\argmax\limits_{t\geq 0} \eta_n{I}_{n}(t)$ when $ \epsilon $ is smaller.

We finally set parameters $ \beta_{nn} = 0.15 $ and $\epsilon=0.001$ and study the effect of parameter $\kappa$ on death-related curves. Figure \ref{fig_2} shows the death-related fractions with different parameter $\kappa$. The parameter $\kappa$ has little effect on non-elderly curves but has large impact on elderly related curves. The time difference $\argmax\limits_{t\geq 0} \eta_e{I}_{e}(t)-\argmax\limits_{t\geq 0} \eta_n{I}_{n}(t)$ is the largest when the parameter $\kappa=3.5$ in three considered parameters $\kappa$. The final elderly deceased fraction $D_e(\infty)$ increases as the parameter $\kappa$.

\begin{figure}[!hbtp]
\begin{center}
\includegraphics[
width=\linewidth
]%
{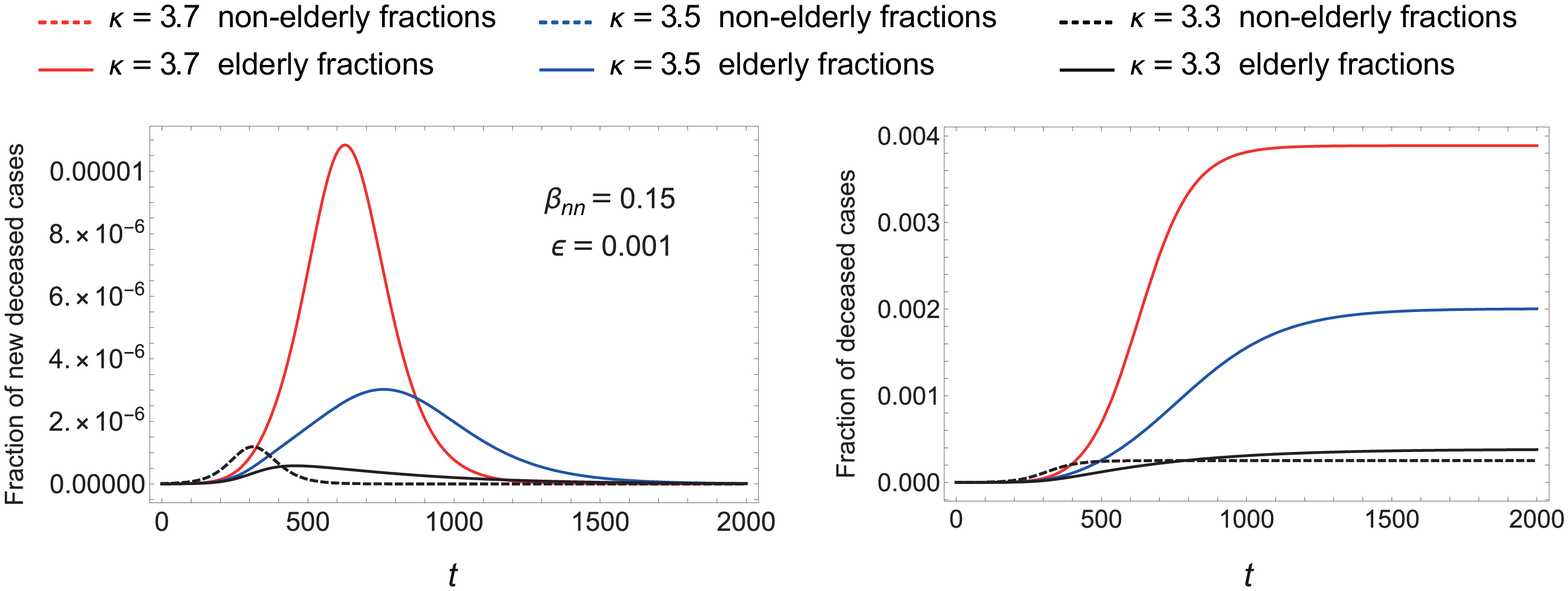}%
\caption{Death-related curves of the two-population SIR model with different parameters $\kappa$. Parameters $ \beta_{nn} $ and $\epsilon$ are set as $ 0.15 $ and $0.001$, respectively. Three different parameters $\kappa$ are considered: $\kappa=3.7$ (red curves), $\kappa=3.5$ (blue curves) and $\kappa=3.3$ (black curves).}%
\label{fig_5a}%
\end{center}
\end{figure}

\begin{figure}[!hbtp]
\begin{center}
\includegraphics[
width=0.8\linewidth
]%
{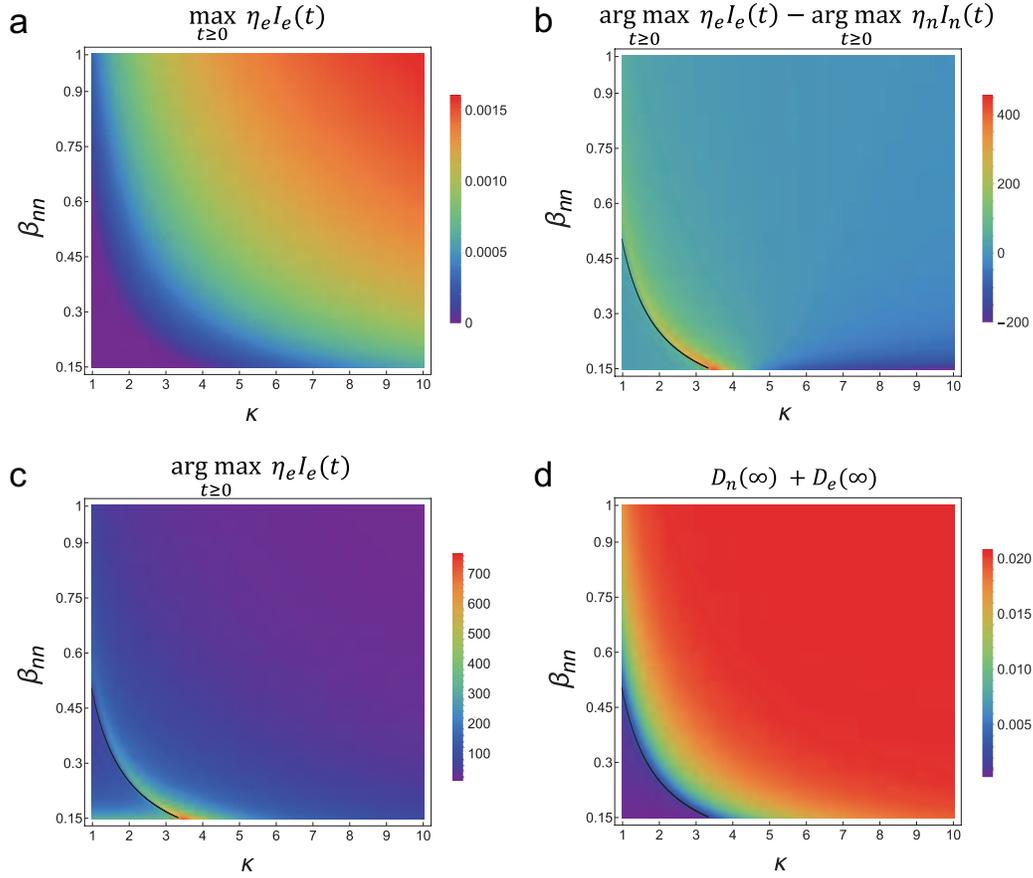}
\caption{Curve features for the two-population SIR model with different parameters $ \beta_{nn} $ and $ \kappa $. The parameter $ \epsilon$ is set to be $ \epsilon=0.001 $. The black curves show the parameters when the infection rate $ \beta_{ee} $ is at the epidemic threshold $ \delta_e/N_e $. The time difference $\argmax\limits_{t\geq 0} \eta_e{I}_{e}(t)-\argmax\limits_{t\geq 0} \eta_n{I}_{n}(t)$ will be large when the infection rate $ \beta_{ee} $ is slightly above the epidemic threshold. The fraction of total deceased individuals will be small when the the infection rate $ \beta_{ee}< \delta_e/N_e$.}
\label{fig_3}
\end{center}
\end{figure}

To better understand the effect of parameters $\beta_{nn}$ and $\kappa$ on death-related curves, we plot the heatmaps as shown in Fig.~\ref{fig_3}. It indicates that there are large time difference $\argmax\limits_{t\geq 0} \eta_e{I}_{e}(t)-\argmax\limits_{t\geq 0} \eta_n{I}_{n}(t)$ when the infection rate $\beta_{ee}$ is around the epidemic threshold. Specifically, suppose that the infection rate between two populations $\beta_{en}\rightarrow 0$, the epidemic threshold for elderly individuals is $ \beta_{ee}= \delta_e/N_e$ (shown as the black curves in Fig.~\ref{fig_3}). The mortality cannot be significantly reduced by only reducing the infection rate among non-elderly individuals $\beta_{nn}$, e.g., closing schools, given that the infection rate $\beta_{ee}$ is above the epidemic threshold. The only efficient way to well reduce the mortality in the two-population SIR model is to keep the infection rate $ \beta_{ee}$ among elderly individuals below the epidemic threshold.

In conclusion, we observe the following interesting curve properties: 1) the death-related curves for non-elderly individuals $\eta_nI_n(t)$ are mainly affected by the infection rate $ \beta_{nn} $, 2) the time difference $\argmax\limits_{t\geq 0} \eta_e{I}_{e}(t)-\argmax\limits_{t\geq 0} \eta_n{I}_{n}(t)$ will be large if the inter-population infection rates $ \beta_{ne} $ and $ \beta_{en} $ are small and the infection rate $ \beta_{ee} $ is slightly above the epidemic threshold, 3) the fraction of eventually deceased cases $D_n(\infty)+D_e(\infty)$ will be small if the infection rate among elderly individuals $ \beta_{ee}< \delta_e/N_e$, 4) only reducing the infection rates among non-elderly individuals cannot significantly reduce mortality. 

The above observations are theoretically explained in Appendix \ref{appendix:appro_theoritical}.

Although mortality can be well reduced by reducing the infection rates among elderly individuals $ \beta_{ee} $, this strategy is not realistic since elderly people necessitate a sufficient amount of social interaction. This work discusses possible strategies to reduce mortality considering the social needs of all the people. Elderly people reduce their social connections with non-elderly individuals and increase their interactions with elderly relationships. Thus their interaction frequency \cite{simmel1890sociale}, which is the total number of social interactions per unit time, is unchanged. We study the effect of reducing connections between elderly and non-elderly individuals on mortality reduction by comparing the mortality in the standard SIR model and the two-population SIR model. To keep the interaction frequency in the standard SIR model and the two-population SIR model to be at the same level, the equivalent infection rate in the standard SIR model is
\begin{equation}
\beta=\beta_{nn}N_n^2+\beta_{ee}N_e^2+(\beta_{ne}+\beta_{en})N_nN_e.
\end{equation}
It holds that $ \beta=\beta_{nn}N_n=\beta_{ee}N_e $ when $ \beta_{ne}\rightarrow 0$, $\beta_{en}\rightarrow 0 $ and $ \beta_{nn}N_n=\beta_{ee}N_e $. Figure \ref{fig_5}a indicates that the fractions of the final deceased individuals for the standard SIR model and the two-population SIR model are the same. The effect of reducing the connection between elderly and non-elderly groups is only to delay the deceased curve, but not to effectively reduce mortality.

\section{Merged SIR model to reduce mortality}
\begin{figure}[!hbtp]
\begin{center}
\includegraphics[
width=\linewidth
]%
{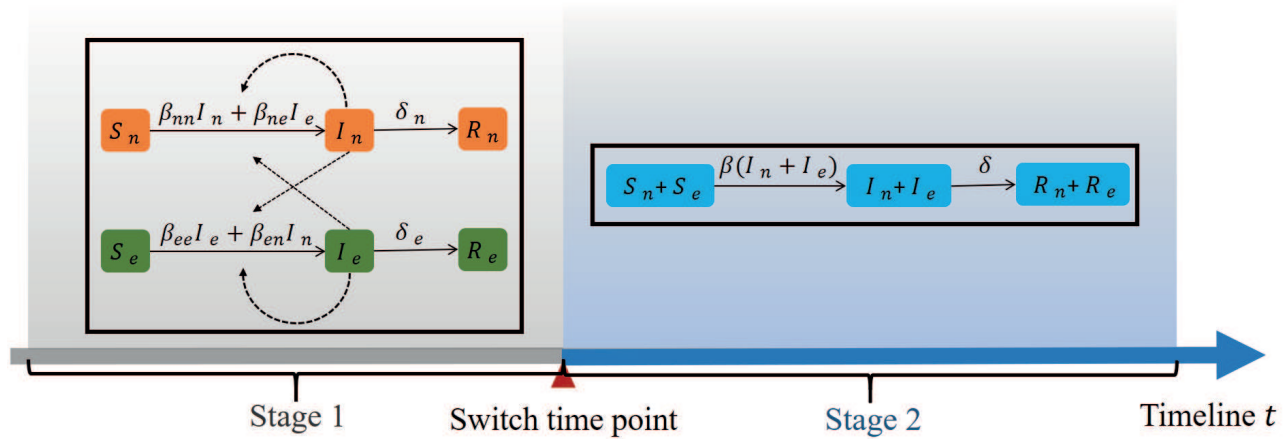}%
\caption{Schematic depiction of the merged SIR model. This model has two stages: the first stage follows the two-population SIR model and the second stage follows the standard SIR model. The physical meaning of this model is to reduce the connection between the elderly and non-elderly populations initially and reconnect these two populations after many non-elderly infected individuals have been recovered. There is a switch time point between these two stages.}%
\label{fig_4}%
\end{center}
\end{figure}

\begin{figure}[!hbtp]
\begin{center}
\includegraphics[
width=\linewidth
]%
{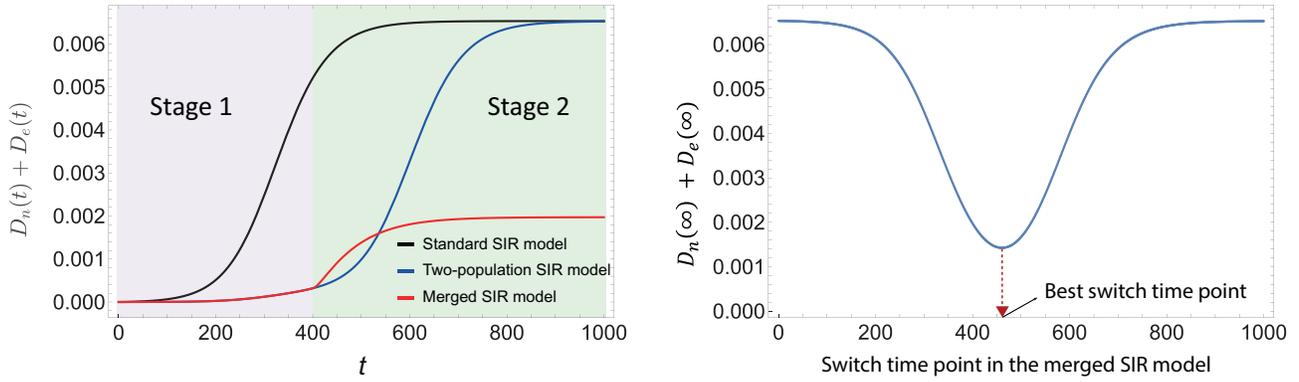}%
\caption{The fractions of deceased individuals in the standard SIR model, the two-population SIR model and the merged SIR model. The parameters for the two-population SIR model are $ \beta_{nn}=0.15 $, $ \epsilon=0.0001 $ and $ \kappa=4 $. We set the infection rate $ \beta=0.12 $ for the standard SIR model to keep the interaction frequency the same among models. The left figure reveals that the two-population SIR model cannot, but the merged SIR model can efficiently reduce mortality. The right figure shows the final deceased fractions with different switch time points, indicating that there is the best switch time point to minimize the final mortality.}%
\label{fig_5}%
\end{center}
\end{figure}

\begin{figure}[!hbtp]
\begin{center}
\includegraphics[
width=\linewidth
]%
{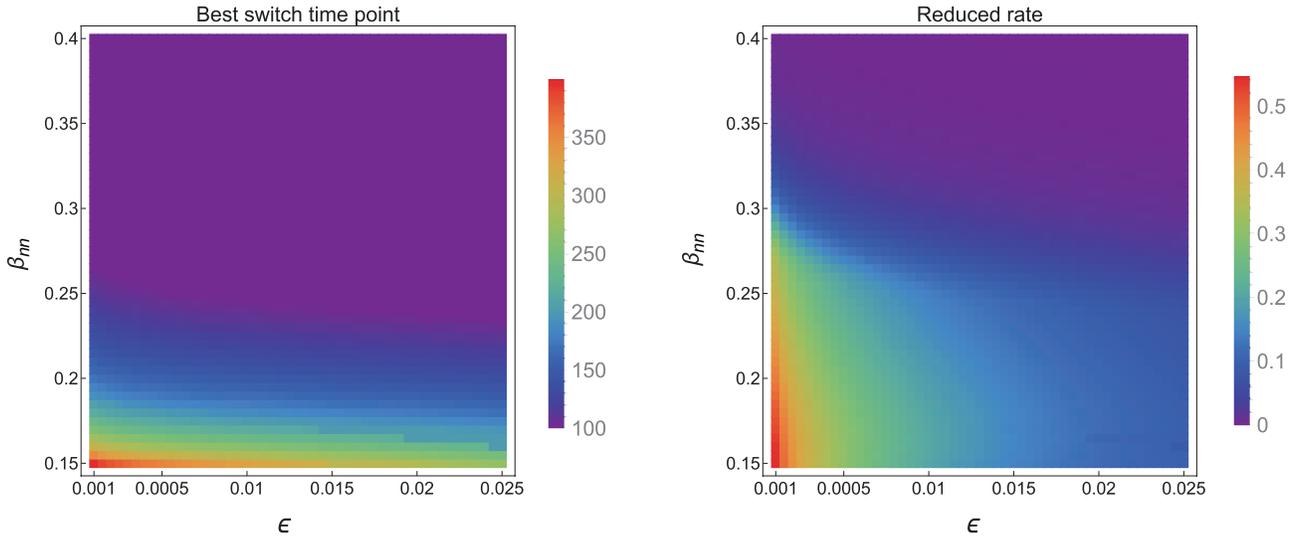}%
\caption{The best switch time point and reduced rate of the final mortality. We choose different parameters $ \beta_{nn} $ and $ \epsilon $ and a fixed parameter $ \kappa=4 $ for the two-population SIR model. The best switch time point will be larger and the final mortality will be reduced more significantly if parameters $ \beta_{nn} $ and $ \epsilon $ are smaller.}%
\label{fig_5b}%
\end{center}
\end{figure}

To effectively reduce mortality, we propose a merged SIR model in which the epidemic spreading follows the two-population SIR model in the first stage and follows the standard SIR model in the second stage. The illustration of the merged SIR model is shown in Fig.~\ref{fig_4}. The reduction of the connection between two populations can delay the pandemic among elderly people. The reconnection of these two populations further protect elderly people due to the herd immunity effect of recovered non-elderly individuals. Figure \ref{fig_5} shows that the merged SIR model can significantly reduce the final deceased fractions and there is the best switch time point to minimize the final mortality. Heatmaps in Fig.~\ref{fig_5b} show the effect of parameters $\beta_{nn}$ and $\epsilon$ on the best switch time point and reduced rate of the final mortality. The reduced rate of the final mortality is defined as 
\begin{equation}
\dfrac{D_e(\infty)+D_n(\infty)-\widetilde{D}_e(\infty)-\widetilde{D}_n(\infty)}{D_e(\infty)+D_n(\infty)},
\end{equation}
where $ D_e(\infty) $ and $ D_n(\infty) $ are respectively the elderly and non-elderly mortality for the two-population SIR model and $ \widetilde{D}_e(\infty) $ and $ \widetilde{D}_n(\infty) $ are respectively the elderly and non-elderly mortality for the merged SIR model. Figure \ref{fig_5b} reveals that the first stage (reducing the connection between non-elderly and elderly people) should take a longer time if parameters $ \beta_{nn} $ and $ \epsilon $ are smaller. 
Besides, the final mortality can be reduced more significantly for smaller parameters $ \beta_{nn} $ and $ \epsilon $.

\section{Two-population SIR epidemic on large complex networks}
\begin{figure}[!hbtp]
\begin{center}
\includegraphics[
width=\linewidth
]%
{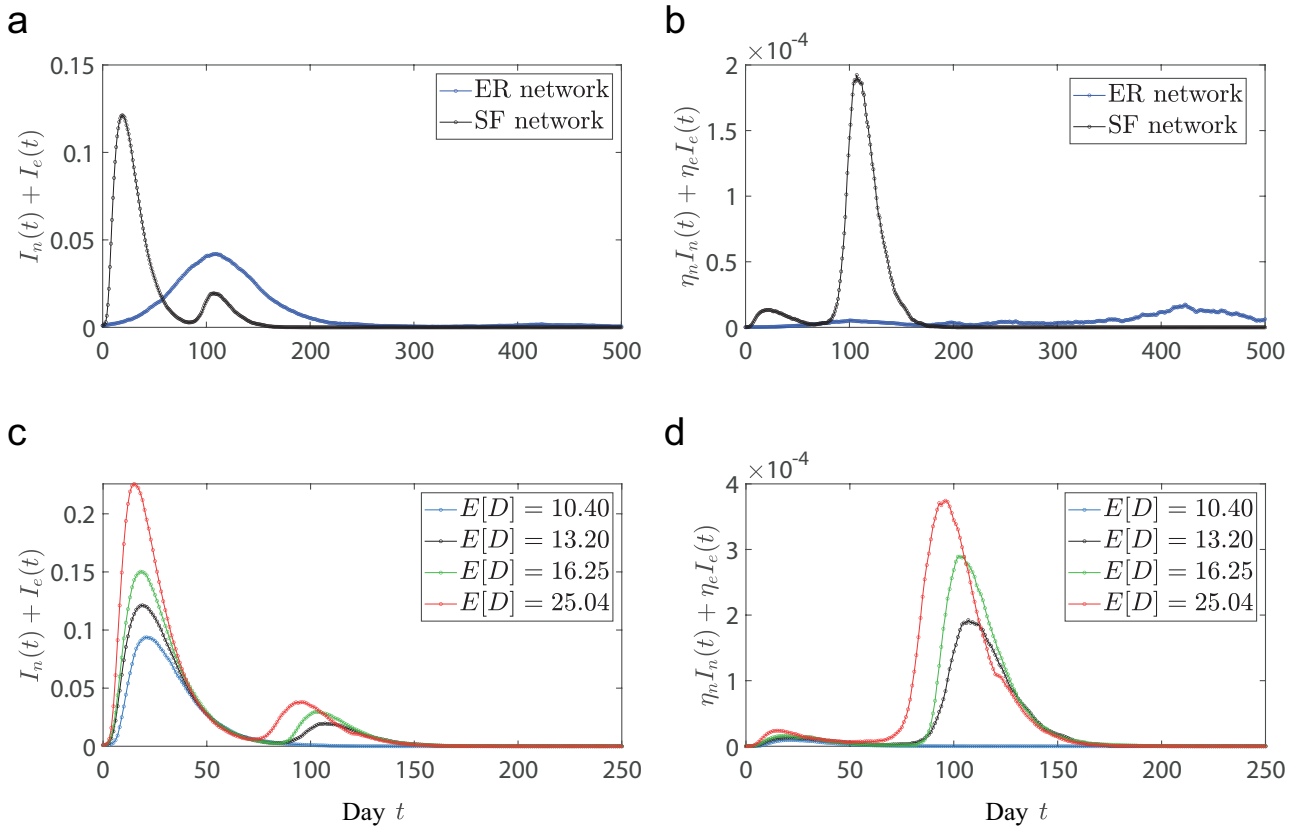}%
\caption{Fractions of infectious cases $ I_n(t)+I_e(t) $ and daily deceased cases $ \eta_nI_n(t)+\eta_eI_e(t) $ for the two-population SIR epidemic on the scale-free network and the Erdős–Rényi network with the network size $ N=10^5 $. The infection parameters are set to be $ \beta_{nn}=0.015 $, $ \kappa=4 $ and $ \epsilon=0.001 $. The spreading in the scale-free network is much faster than the spreading in the Erdős–Rényi network. Figures c and d show the effect of mean degree $ E[D] $ of the scale-free network on the two-population SIR epidemic. With the increase of the scale-free networks' link density, there are more individuals infected and deceased. The exponent in the scale free networks is set as $ \gamma=2.5 $. The minimum degree of the scale-free network is set to be $ 5 $.}%
\label{fig_6}%
\end{center}
\end{figure}

We apply the Monte Carlo method \cite{juher2009analysis} to simulate the two-group SIR epidemic on complex networks. In this work, we consider large networks with network size $ N=10^5 $ generated by the configuration model \cite{newman2003structure} and the simulation starts from $ 100 $ non-elderly infected individuals. We first compare the simulation results on the scale-free network and the Erdős–Rényi random network to analyze the effect of network heterogeneity on epidemic curves. The network size $ N $ and mean degree $ E[D] $ of the Erdős–Rényi random network are the same as the scale-free network. Figure \ref{fig_6}a and Fig.~\ref{fig_6}b indicate that the epidemic spreading in the scale-free network is much faster than the spreading in the Erdős–Rényi random network due to the super spreaders. Figure \ref{fig_6}c and Fig.~\ref{fig_6}d illustrate that the epidemic spreads quicker when the mean degree $E[D]$ is higher.

\begin{figure}[!hbtp]
\begin{center}
\includegraphics[
width=\linewidth
]%
{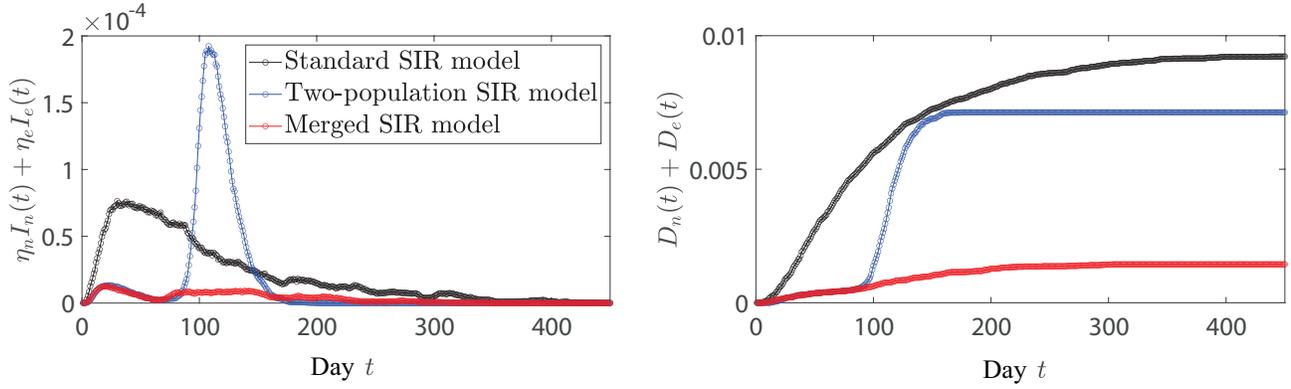}%
\caption{Fractions of daily deceased cases $ \eta_nI_n(t)+\eta_eI_e(t) $ and deceased cases $ D_n(t)+D_e(t) $ for the standard SIR epidemic, the two-population SIR epidemic and the merged SIR epidemic on the scale-free network with the network size $ N=10^5 $. The infection rate in the standard SIR model is $ \beta=0.012 $. The infection rates in the two-population SIR model are the same as Fig.~\ref{fig_6}. Different from the result in Fig.~\ref{fig_5}, the final deceased fraction for the two-population SIR model is lower than the standard SIR model. The final deceased fraction for the merged SIR model is the smallest, indicating that the merged SIR model is the best strategy to reduce the mortality.}%
\label{fig_7}%
\end{center}
\end{figure}

We simulate the standard SIR model, the two-population SIR model and the merged SIR model on the scale-free network as shown in Fig.~\ref{fig_7}. Different from the results as demonstrated in Fig.~\ref{fig_5}, for the epidemic spreading on complex networks, the final mortality for the two-population SIR model is lower than the standard SIR model since a part of susceptible elderly people can be protected by their recovered non-elderly relationships. This type of local immunity, which differs from the herd immunity, can only be observed in the epidemic spreading on networks. The merged SIR model is the best strategy to reduce mortality.

\begin{figure}[!hbtp]
\begin{center}
\includegraphics[
width=\linewidth
]%
{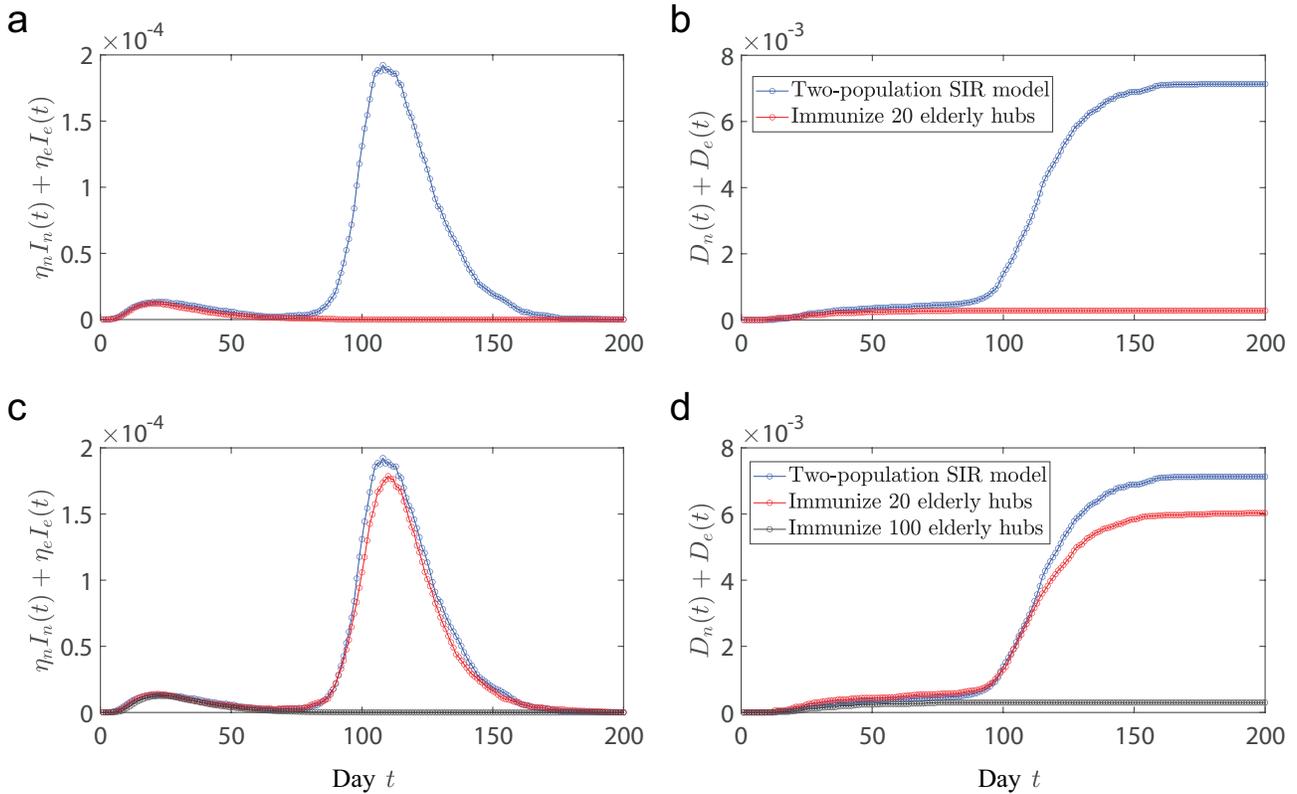}%
\caption{Effect of immunizing rare elderly hub individuals on reducing the final mortality. Figures a and b respectively show the fractions of daily deceased cases $ \eta_nI_n(t)+\eta_eI_e(t) $ and deceased cases $ D_n(t)+D_e(t) $ for the two-population SIR epidemic with and without immunizing elderly hub individuals. We immunize $ 20 $ elderly individuals with the largest degree in the simulation of the two-population SIR model on the scale-free network with $ 10^5 $ population assuming that the vaccines are $ 100\% $ effective. The mortality can be significantly reduced by immunizing such rare specific individuals. Figures c and d show the fractions when the vaccines are $ 80\% $ effective. It requires more vaccine doses to effectively reduce the mortality if the vaccines are less effective.}%
\label{fig_8}%
\end{center}
\end{figure}

Given that there have been COVID-19 vaccines but the vaccine is still insufficient, it is valuable to study the strategy to reduce mortality by immunizing specific population. There are rare elderly hub individuals in social networks, e.g., the priests, which are the virus's primary route of transmission from non-elderly to elderly people. Figure \ref{fig_8}a and Fig.~\ref{fig_8}b reveal that the final mortality can be significantly reduced by only immunize $ 20 $ elderly hub individuals in $ 10^5 $ population assuming that the vaccines are $ 100\% $ effective. In reality, the COVID-19 vaccine efficacy cannot reach $ 100\% $ and thus we analyze the situation when the vaccines are $ 80\% $ effective. Figure \ref{fig_8}c and Fig.~\ref{fig_8}d illustrate that more elderly hub individuals require to be immunized to reduce mortality efficiently.

\clearpage
\section{Conclusions}
Since early 2020, scientists have found that COVID-19 is substantially more dangerous for the elderly. Elderly people's interactions with their non-elderly relationships are reduced to lower the risk of being infected and deceased. This work applies the two-population SIR model to describe the COVID-19 pandemic when the connections between elderly and non-elderly individuals are significantly reduced. We analyze how the reduction of connections between two populations can affect the COVID-19 pandemic, especially the mortality. It reveals that severing ties between two populations can postpone the pandemic but not effectively cut mortality. We further find that reconnecting two populations at an appropriate time can significantly lessen the final mortality. Assuming that rare vaccines are available, this study recommends immunizing elderly hub individuals first to better decrease mortality.

\bibliography{bibitem}
\bibliographystyle{unsrt}
\appendix
\section{Theoretical explanation of the death-related curve features}\label{appendix:appro_theoritical}
\begin{figure}[!hbtp]
\begin{center}
\includegraphics[
width=0.5\linewidth
]%
{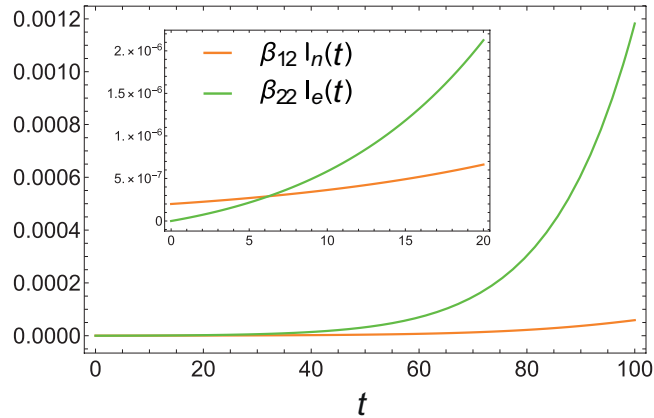}%
\caption{The curves of $ \beta_{ee} I_{e}(t) $ (the green curves) and $ \beta_{ne} I_{n}(t) $ (the orange curves) for the two-population SIR model with parameters $ \beta_{nn}=0.2 $, $ \epsilon=0.01 $ and $ \kappa=4 $. The value $ \beta_{ee} I_{e}$ is much larger than $ \beta_{ne} I_{n} $ after the very initial stage.}%
\label{fig_3b}%
\end{center}
\end{figure}

In Equations \eqref{eqn:well-mixed_sir_model}, we have that
\begin{align}
\dfrac{dI_{n}(t)}{dt} &= \beta_{nn} S_{n}(t)I_{n}(t)+\beta_{en} S_{n}(t)I_{e}(t)-\delta_n{I}_{n}(t) \\
&=(\beta_{nn}I_{n}(t)+\beta_{en}I_{e}(t))S_{n}(t)-\delta_n{I}_{n}(t).
\end{align}
Since non-elderly individuals are the majority in the whole population and the virus spreads from non-elderly individuals, it holds that $ I_{n}(t)\gg I_{e}(t) $ at the initial stage of the spreading. Moreover, the infection rates hold that $ \beta_{nn}\gg\beta_{en}$ and thus we have $ \beta_{nn}I_{n}(t)\gg \beta_{en}I_{e}(t) $, which indicates that elderly infections have little impact on the non-elderly susceptible individuals and the initial infection curve $ I_n(t) $ for non-elderly individuals is close to the result in standard SIR model:
\begin{equation}
\left\{
\begin{aligned}
&\dfrac{dS_{n}(t)}{dt} = -\beta_{nn} S_{n}(t)I_{n}(t),\\
&\dfrac{dI_{n}(t)}{dt} = \beta_{nn} S_{n}(t)I_{n}(t)-\delta_n{I}_{n}(t),\\
&\dfrac{dR_{n}(t)}{dt} = \delta_n{I}_{n}(t).
\end{aligned}
\right.
\label{eqn:well-mixed_sir_model_non_elderly}
\end{equation}
This explains why curve features for non-elderly individuals are little affected by parameters $ \epsilon $ and $ \kappa $. When time $ t\rightarrow 0 $ and the inter-population infection rate $ \beta_{ne}\rightarrow 0 $, the fraction for non-elderly infectious individuals is close to the exponential function
\begin{equation}
I_n(t)\approx I_n(0)e^{(\beta_{nn}S_n(0)-\delta_n)t}.
\label{eqn:sir_model_non_elderly_expo}
\end{equation} 
Substitute \eqref{eqn:sir_model_non_elderly_expo} into the equation $ {dI_{e}(t)}/{dt} = \beta_{ne} S_{e}(t)I_{n}(t)+\beta_{ee} S_{e}(t)I_{e}(t)-\delta_e{I}_{e}(t) $ in \eqref{eqn:well-mixed_sir_model} and we have that,
\begin{equation}
\begin{aligned}
\dfrac{dI_{e}(t)}{dt} &\approx \beta_{ne}S_e(t)I_n(0)e^{(\beta_{nn}S_n(0)-\delta_n)t}+\beta_{ee}S_e(t)I_e(t)-\delta_e I_e(t)\\
&\approx \beta_{ne}S_e(0)I_n(0)e^{(\beta_{nn}S_n(0)-\delta_n)t}+(\beta_{ee}S_e(0)-\delta_e)I_e.
\end{aligned}
\end{equation}
We simplify the above equation by letting $ a=\beta_{ne}S_e(0)I_n(0) $, $ b=\beta_{ee}S_e(0)-\delta_e $ and $ m=\beta_{nn}S_n(0)-\delta_n $: 
\begin{equation}
\dfrac{dI_{e}}{dt} \approx ae^{mt}+bI_e(t).
\end{equation}
By solving the above equation and combining the fact that $ I_e(0)=0 $, the fraction of elderly infectious individuals when time $ t\rightarrow 0 $ is
\begin{equation}
I_e(t)\approx\dfrac{a(e^{mt}-e^{bt})}{m-b}.
\label{eqn:sir_model_elderly_appro}
\end{equation}
This equation is the difference of two exponential functions, indicating that the initial curve for elderly individuals cannot be well described by an independent SIR model. Besides, at the initial stage of spreading, the growth rate of $ I_e(t) $ decreases with the deceasing of $ \beta_{ne} $. A slower growth of $ I_e(t) $ at the initial stage will delay the further curve and peak position. Figure \ref{fig_3b} shows the values $ \beta_{ee} I_{e}(t) $ and $ \beta_{ne} I_{n}(t) $ in equation $ {dI_{e}(t)}/{dt} = \beta_{ne} S_{e}(t)I_{n}(t)+\beta_{ee} S_{e}(t)I_{e}(t)-\delta_e{I}_{e}(t) $. It reveals that the value $ \beta_{ne} I_{n}(t) $ dominates only at the very initial stage and the value $ \beta_{ee} I_{e}(t) $ dominates the later stage. When the infection rate between two groups is relatively small, the later curve for elderly people will be close to the independent SIR model:
\begin{equation}
\left\{
\begin{aligned}
&\dfrac{dS_{e}(t)}{dt} = -\beta_{ee} S_{e}(t)I_{e}(t),\\
&\dfrac{dI_{e}(t)}{dt} = \beta_{ee} S_{e}(t)I_{e}(t)-\delta_e{I}_{e}(t),\\
&\dfrac{dR_{e}(t)}{dt} = \delta_n{I}_{e}(t).
\end{aligned}
\right.
\label{eqn:well-mixed_sir_model_non_elderly}
\end{equation}

Robert Schaback \cite{schaback2020covid} proved that, for the independent SIR model, when the initial value $ S_e(0) \approx N_e$ and $ S_e(0)\beta_{ee}>\delta_e $, the upper bound of the peak position for the fraction of infectious individuals is
\begin{equation}
\dfrac{\delta_e}{I_e(0)\beta_{ee}}\log\left(\dfrac{S_e(0)\beta_{ee}}{\delta_e}\right).
\end{equation}
The largest peak position can be obtained when $ \beta_{ee} $ is slightly larger than $ \delta_e/N_e(0)$.

\end{document}